\documentclass[%
 reprint,
 amsmath,amssymb,
 aps,
]{revtex4-2}

\usepackage{natbib}

\usepackage{orcidlink}

\usepackage{color}
\usepackage{graphicx}
\usepackage{dcolumn}
\usepackage{bm}
\usepackage[pagewise,columnwise]{lineno}
\usepackage{orcidlink}

\begin{document}

\preprint{APS/123-QED}

\title{Viscosity of non equilibrium hot $\&$ dense QCD drop formed at LHC}

\author{J. R. Alvarado Garc\'ia\orcidlink{0000-0002-5038-1337}} 
\email{j.ricardo.alvarado@cern.ch}
\author{I. Bautista\orcidlink{0000-0001-5873-3088}}
\email{irbautis@cern.ch}
\author{A. Fern\'andez T\'ellez\orcidlink{0000-0003-0152-4220}}
\affiliation{
Facultad de Ciencias F\'isico Matem\'aticas, Benem\'erita Universidad Aut\'onoma de Puebla, 1152, Puebla 72570, M\'exico}
\author{P. Fierro\orcidlink{0000-0002-3206-301X}}
\affiliation{
Instituto de F\'isica Luis Rivera Terrazas, Benem\'erita Universidad Aut\'onoma de Puebla, (IFUAP), Apartado postal J-48 72570 Puebla. Pue., M\'exico
}%

%



\begin{abstract}
{
We compute the bulk, $\zeta$, and shear, $\eta$, viscosity over entropy density, $s$, for the QCD matter formed in small collision systems at LHC. We consider a scenario of the String Percolation Model by proposing a global form of the color reduction factor that describes both the thermodynamic limit and its maximum deviation due to small-bounded effects. Our method involves estimations at vanishing baryon-chemical potential, assuming local equilibrium for string clusters in the initial state. To compute $\eta/s$, we employed a kinetic approach that accounts QCD states as an ideal gas of partons, while $\zeta/s$ is computed by using two different approaches: a simple kinetic formula and the causal dissipative relativistic fluid dynamics formulation. Our results align with Lattice QCD computations and Bayesian methods and are consistent with holographic conjecture bounds. Furthermore, our findings support the notion of a strongly interacting medium, similar to that observed in nuclear collisions, albeit with a phase transition occurring outside the thermodynamic limit.

}

\end{abstract}

\maketitle





\section{Introduction}

The String Percolation Model (SPM) has described successfully the collective effects on medium formed at heavy ion collisions from the Relativistic Heavy Ion Collider (RHIC) to the Large Hadron Collider's (LHC) energies \cite{Bautista:2011mc,Bautista:2012fz,Bautista:2013ema,Andres:2016mla,Srivastava:2011vz,Scharenberg:2010zz,Ramirez:2020vne,Braun:1999hv,Braun:2015eoa}. At these energy regimes non-perturbative QCD takes a major role in describing phenomena through phenomenological models, such as the characteristics of the QCD phase diagram, and the phase transition properties, that can be studied in terms of the systems' thermodynamic quantities,  transport coefficients, and bulk properties \cite{Braun:2015eoa}. 
Recently, to obtain the values of bulk and shear viscosity from nuclear collisions, relativistic hydrodynamics have been used to calculate the temperature dependence of these coefficients
\cite{Roy:2012np,Parkkila:2021tqq,JETSCAPE:2020mzn,Parkkila:2021tqq,Niemi:2015qia,Niemi:2015voa,Bernhard:2019bmu}, as well as theoretical limits in AdS/CFT correspondence and holographic link with Quark-Gluon Plasma (QGP) \cite{Policastro:2001yc,Buchel:2007mf}.

The String Percolation Model uses the percolation theory for non-perturbative heavy ion physics, by utilizing as main objects the effective color sources and describing their physical properties such as color field, momentum, and multiplicity \cite{Braun:1999hv,Braun:2015eoa}. The functional form of its main parameters was deduced from Monte Carlo simulations considering thermodynamic limit. And more recently, the results of clustering of color sources that consider finite size, profile distribution function, and the initial geometry effects (which correspond to systems far from the thermodynamic limit, that from now on we will denote as nonTL) were studied and compared with what was previously reported for known observables, such as the thermal-like temperature extracted from the transverse momentum spectra and the estimation of thermodynamic quantities \cite{Ramirez:2017oef,Ramirez:2019ekt}. 
In this work, we study from a phenomenological view, the signatures of collective effects reported on \cite{Bala:2016hlf,Loizides:2016tew}. Specifically, we studied the bulk properties like the modification of the behavior of the speed of sound and we calculated the bulk viscosity coefficient; which are significant for non-thermal equilibrium systems and highlights the benefits that our approach provides. With this new perspective on the SPM, we can see new light on how the effects of bulk properties weigh on reaching critical temperature on nonTL systems, which is the first step for computing the bulk properties for non-thermal equilibrium systems.

In the following section, we present the basics of the String Percolation Model. Then,  we proceed with exploring the consequences of considering the nonTL scenarios in the SPM framework in Sec.~\ref{chIII}. In Secs.~\ref{chIV} and \ref{chV} we introduce the temperature and energy density respectively as usually reported. Finally, in Secs.~\ref{chVI} and~\ref{chVII} we discuss the results on bulk and shear viscosity to entropy density ratios of TL and nonTL scenarios in the SPM. 

\section{Percolation color sources}
\label{chII}
The string percolation model uses percolation theory which is closely associated with the study of phase transitions and transport phenomena  \cite{Stauffer:1978kr,Kirkpatrick:1973ni,Broadbent:1957rm}, necessary to characterize the medium formed in ultra-relativistic heavy ion collisions.
The interaction between colliding nuclei is effectively represented by the formation of extended color flux tubes which are stretching among the colliding partons. We consider the projection of the color flux tubes in the transverse plane, from now on named strings, in order to use the 2-dimensional percolation approach, which unlike in the thermodynamic approaches is able to describe a phase transition without defining a temperature in a thermodynamic equilibrium \cite{Braun:1999hv,Braun:2015eoa}. 

The strings can be seen as small disks characterized by their transverse area which is on average taken as $S_{0} = \pi r_0^2$ ($\sim 3.5$ mb from parton-parton cross section of bilocal correlation functions \cite{Amelin:1993cs,string} and $r_0$ the radius of a single disk). 

The percolation approach is only used to characterize the string density in the initial state. Based on the initial density this approach gives the probability of forming a spanning color strings system which means that in a 2-dimensional percolation approach a geometrical connected system of disks represents a connected color flux tubes state. The spanning system will then later evolve due to the Schwinger mechanism where a temperature is defined as the corresponding thermal $T$-slope of $p_T$-exponential momentum distribution. In this sense, the temperature is associated with the final state of experimental observations. As we will detail in Sec.~\ref{chIV}.

\begin{figure}[ht!]
    \centering
    \includegraphics[scale=0.45]{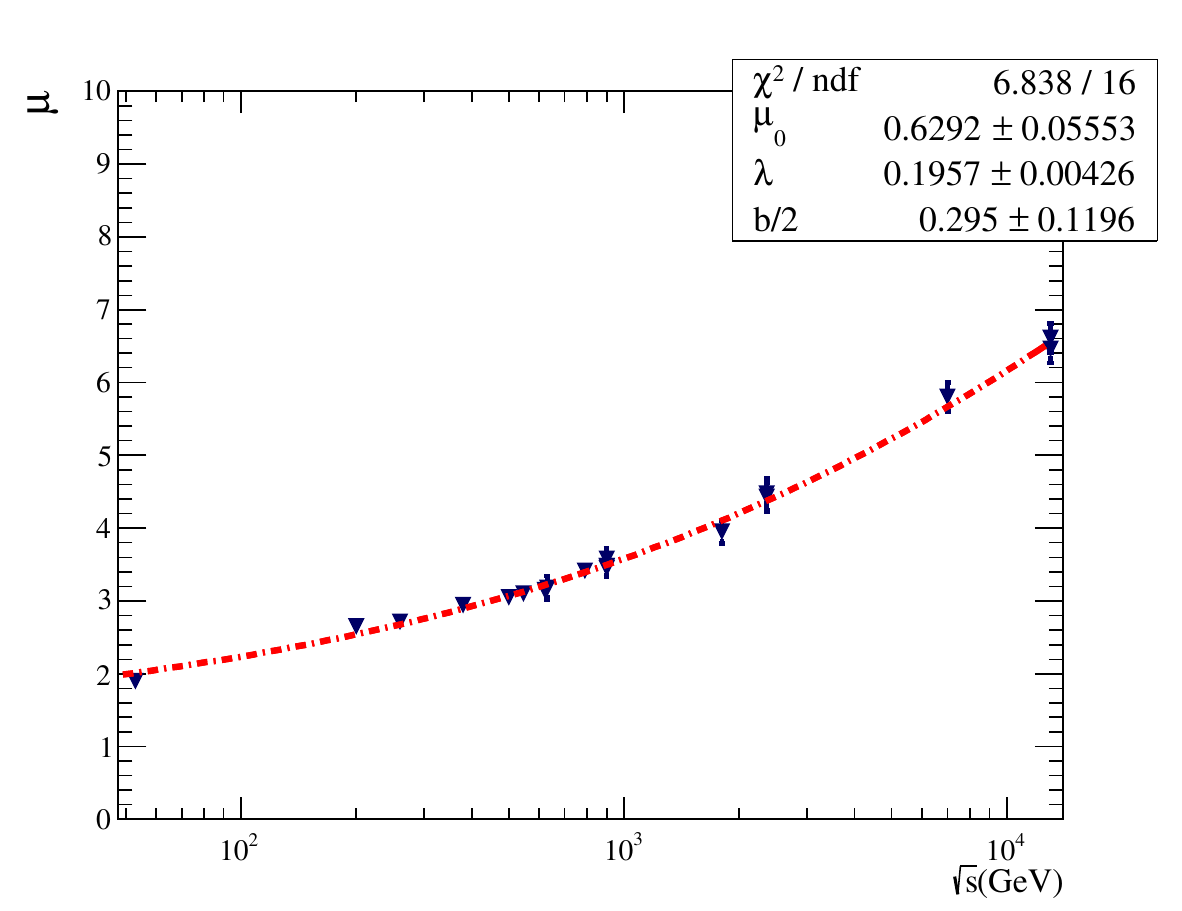}
    \caption{Fit over experimental data for multiplicity measured on $pp$ collisions from 53 MeV to 13 TeV, data taken from Refs. \cite{UA1:1989bou,UA5:1987rzq,STAR:2008med,CDF:1989nkn,ALICE:2010cin,CMS:2010wcx,CMS:2010tjh,ALICE:2015qqj,CMS:2015zrm}.}
    \label{fig:multi}
\end{figure}

For characterizing the system, an order parameter is introduced, which depends on the area fraction occupied by a determined number of strings:
\begin{equation}
    \xi = \frac{S_0}{S}  N_s,
    \label{eq.stringd}
\end{equation}
where $N_{s}$ is the number of initial strings in an event, which for a minimum bias distribution escalates with energy as a power law \cite{Bautista:2015kwa}:  
\begin{equation} N_{s} = 2 + 4  \frac{S_0}{S}  \Bigg( \frac{ \sqrt{s} }{ m_p } \Bigg)^{2 \lambda},
\end{equation} 
with $m_p$ the proton mass and $\lambda = 0.196 \pm 0.005$ a fit parameter shown in Fig.~\ref{fig:multi}.
For a large number of strings in an event, it is required to have a large number of partonic interactions which can be achieved at high collision energies or a large number of colliding partons, like $AA$ collisions. The created disk's distribution and overlapping (cluster formation) marks a phase transition when the system starts to percolate for a critical value of the string density $\xi_c$ \cite{Braun:2015eoa}, which depends on the characteristics of the system \cite{Rodrigues:1998it}. 

In $pp$, collisions the areas $S$ and $S_0$  can be described in terms of the radii $r_0 \simeq 0.2385$ fm \cite{Bialas:1999zg,DiGiacomo:1992hhp,DiGiacomo:1993jt,Bali:1994de} and $R_p \simeq 1$ fm (the radius of a proton). 
However, to have a more precise description of the overlapping area, we define it as an ellipse in terms of an effective impact parameter $b$:
\begin{equation}
    S
    =\pi \left( R_p - \frac{b}{2}\right) 
    \sqrt{R_p^2 - \left(\frac{b}{2}\right)^2}. 
    \label{eq.3}
\end{equation}
For greatly overlapped areas we can approximate Eq.~\eqref{eq.3} as the area of a circle $S \simeq  \pi R_p^2$.

Cluster formation implies the creation of new color sources in which color fields are the vector sum of the overlapped areas' color fields. Due to the random orientation of the color fields, the mixed terms vanish. Thus the color field intensity is proportional to the squared color charges of the original strings $\sqrt{n}$, and, in consequence, multiplicity $\mu$ is: 

\begin{equation}
\frac{\mu}{\mu_0} 
= \frac{\langle \sqrt{n}\text{ } \rangle}{S_0} S
= N\frac{\langle \sqrt{n} \text{ } \rangle}{\xi} , 
\label{eq:15}
\end{equation}
where $\mu_0$ is the multiplicity of a single string \cite{Braun:1999hv,Braun:2015eoa}. So, the number of charged particles generated in the midrapidity region is directly proportional to the initial number of strings of the system \cite{Bautista:2015kwa}:
\begin{equation}
\mu = \mu_0 F( \xi )N_s ,
\end{equation}
where $\mu_0 \sim 0.63$ is calculated from fit over data \cite{UA1:1989bou,UA5:1987rzq,STAR:2008med,CDF:1989nkn,ALICE:2010cin,CMS:2010wcx,CMS:2010tjh,ALICE:2015qqj,CMS:2015zrm} shown in Fig.~\ref{fig:multi}.

From the above equation, a geometric scaling function appears, namely the Color Reduction Factor $F( \xi )$, which emerges naturally from cluster formation \cite{Braun:1999hv,Braun:2015eoa}. This function increases with the string tension of the cluster and the average momentum fraction of the partons $\langle p_{T}^{2} \rangle$. 
In the thermodynamic limit, $F(\xi)$ depends on the string density $\xi$ as \cite{Braun:1999hv,Braun:2015eoa,DiasdeDeus:2006xk}:
\begin{equation}
    F( \xi ) = \sqrt{ \frac{1-e^{- \xi} }{ \xi } } .
\label{eq6}
\end{equation}
This geometric scaling function describes the universality of the scaling law of the system, in correspondence to heavy nuclei.

\section{Non-thermodynamic limit color reduction factor}
\label{chIII}

In the field of heavy-ion collisions, understanding the behavior of small collision systems is of paramount importance. Traditional studies often make predictions for these systems by obtaining $F(\xi)$ by fitting experimental measurements and assuming TL \cite{Bautista:2015kwa,Gutay:2015cba,Scharenberg:2018oyj,Mishra:2022kre}. While useful, this approach may not be the best method for describing systems far from reaching the TL. To address this, we consider the impact of system size effects \cite{BautistaGuzman:2017ubi} and other initial state conditions \cite{Ramirez:2017oef,Ramirez:2019ekt,Garcia:2022ozz}. 

To account for maximum deviations from the TL in $F(\xi)$, we propose a universal function for the color reduction factor that includes an additional damping term. This function captures the TL limit and fits deviations from the geometric scaling function as revealed in simulation results that consider small-bounded effects. This factorization can be expressed as:

\begin{equation}
\begin{split}
    F_s(\xi) &= m\sqrt{\frac{1-\exp(-\xi)}{\xi}}+c\sqrt{\frac{1+\exp(-\xi)}{\xi}},
    \end{split}
\label{eq7}\end{equation}

where $m=0.7714731 \pm 0.01468$ is the weight parameter of the TL contribution to $F(\xi)$ (the typical percolation model), and $c=0.0609589 \pm 0.007527$ is the weight parameter of the deviation from the nonTL in the percolating system. The difference $\Delta F = F(\xi) - F_s (\xi)$ represents the strength of the fluctuations of the percolating object's properties. The equation reduces to Eq.~\eqref{eq6} when $c=0$ (indicating no additional damping from finite size effects) and $m=1$ (representing the TL). 

This new formulation allows a larger suppression effect above critical string density, which is significant even when $c$ is small due to the large deviation from TL exhibited by the area covered by disks included in our fit \cite{Ramirez:2019ekt}. It also explains similar deviations reported for regions just below the critical string density \cite{Scharenberg:2018ocw,Scharenberg:2018oyj}. 

With this improvement to the SPM, we can now better account for effects in a broader range of multiplicity experimental data \cite{STAR:2003fka,ALICE:2018vuu,ALICE:2013txf,ALICE:2019dfi,ALICE:2015qqj,ATLAS:2016zkp,CMS:2012xvn,CMS:2013pdl,CMS:2017eoq} and improve the model's predictions for transport coefficients and bulk properties for temperature regions below the critical temperature for nonTL systems. We will demonstrate these improvements in subsequent sections.

We also estimated the effective region of $F_s(\xi)$ for systems that lie between the TL and nonTL using a linear interpolation method. For the derived model-dependent observables, the effective region was calculated through an uncertainty propagation method.


\section{Thermal Distribution}
\label{chIV}

As mentioned before, the local effective thermodynamic quantities are connected with the geometrical properties of the percolating system through $F_s(\xi)$. The string density rules the cluster distribution, and in consequence, the behavior of all thermodynamic quantities, such as temperature which involves the Schwinger mechanism for non-massive particles, on which the strings with higher tension $x$ will break producing $q\bar{q}$ and $qq$-$\bar{q}\bar{q}$ pairs which later on will combine producing the final state hadrons.

So, the transverse momentum distribution of charged particles is given by \cite{Schwinger:1962tp}:  
\begin{equation}
\frac{dN}{dp_T^2} \sim \exp\left( -\pi \frac{p_T^2}{x^2 }  \right) .
\label{e8}
\end{equation}

The tension of the string $x^2$ fluctuates around its mean value, $\langle x^2 \rangle$, describing a Gaussian distribution of the fluctuations which convolutes with \eqref{e8} giving a thermal-like distribution characterized by the mean transverse momentum of a single string $ \langle p_T^2 \rangle_0  =  \langle x^2 \rangle  F_s(\xi) / \pi $ \cite{Bialas:1999zg,Braun:2015eoa}:
\begin{equation}
\frac{dN}{dp_T^2} \sim \exp\left( -p_T \sqrt{ \frac{2F_s(\xi)}{\langle p_T^2\rangle_0} } \right) ,
\end{equation}
from where we estimate the temperature in the same way as in the Boltzmann distribution:
\begin{equation}
    T(\xi)=\sqrt{ \frac{\langle p^2_{T} \rangle_{0}}{2F_s(\xi)}}.
\label{eqTemp}
\end{equation}

The critical string density depends on the system's characteristics \cite{Garcia:2022ozz,Ramirez:2019ekt,Ramirez:2017oef}.
We consider the critical temperature in terms of critical string density $\xi_c$ = 1.128 \cite{mertens} in the same way as \cite{DiasdeDeus:2016itj,Bautista:2015kwa}, where $T_c = T(\xi_{c})$, so that, in TL:
\begin{equation}
    \frac{T}{T_c} = \sqrt{\frac{F(\xi_c)}{F(\xi)}} = \frac{0.879947816}{\sqrt{F(\xi)}}.
\end{equation}

Furthermore, we also see a shift in the critical temperature, now reached at lower $\xi$ for $F_s(\xi)$. We consider the critical temperature $T_c = 154\pm 9$ \cite{Bazavov:2011nk}, and its respective deviations with $\xi$ to estimate an effective area of the observables as a function of $T/T_c$.


\section{Energy density}
\label{chV}

\begin{figure}[ht!]
\begin{center}
\includegraphics[scale=0.45]{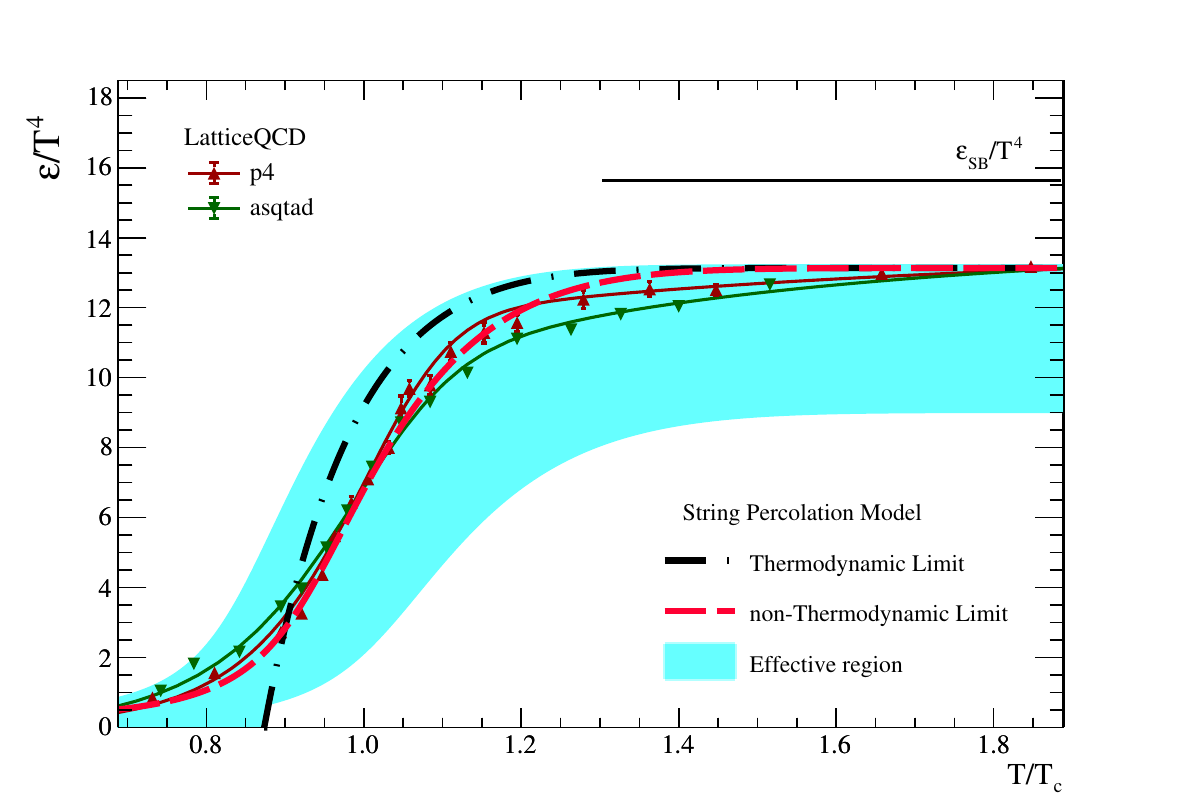}
\caption{
Energy density over $T^4$ behavior with respect to $T/T_c$. The cyan area shows the effective region estimated with $\Delta F$. The red dashed line shows the energy density computed in nonTL. 
While the black dotted-dashed line is the TL. 
\cite{Braun:2015eoa}. We include a comparison with Lattice QCD computations p4 (maroon triangles) and asqtad (green triangles) actions with their respective parametrization (continuum lines) \cite{Bazavov:2009zn}.
}
\label{f1}
\end{center}
\end{figure}

In the Stephan-Boltzmann approximation, quarks and gluons are assumed to be noninteracting and massless \cite{Hagedorn:1983wk}. So that the energy density $\varepsilon$ 
is an order parameter in the phase transition from the Hadron Gas (HG) to QGP, revealing an increment in the internal degrees of freedom.

Moreover, the string density $\xi$ is the local order parameter in the SPM that marks the geometric phase transition \cite{Braun:2015eoa}. 
In references \cite{Braun:2015eoa,DiasdeDeus:2012uc,Sahoo:2017umy}, energy density from the Bjorken boost invariant 1D hydrodynamics formula \cite{Bjorken:1982qr} is found to be proportional to $\xi$. 
Given that each initial state string can be interpreted as the extended fields among the interacting partons, which has a direct contribution to energy density.  
This key idea holds for small collision systems \cite{Bautista:2015kwa,Gutay:2015cba}. Consequently, we use the following relation to estimate energy density:
%
\begin{equation}
\varepsilon = \varsigma \xi,
\label{Benergy}
\end{equation} 
where we found in the same way as \cite{DiasdeDeus:2012uc} that $\varsigma= \varepsilon_c /\xi_c  = 0.5601$ $\text{GeV/fm}^3$, and see a shift in the critical temperature $T_c(F_s)/T_c(F) =1.096$ with respect to the TL scenario.

As we can see in Fig.~\ref{f1}, the behavior of the energy density over $T^4$ as a function of $T/T_c$ agrees with Lattice's calculations using staggered fermion actions p4 and asqtad \cite{Bazavov:2009zn}.  The observed increment of energy density is related to a rise in the number of degrees of freedom from the Hadron-Gas phase where there are fewer than in the QGP phase, and the quantum color numbers contribute to the energy density \cite{Karsch:2001vs}.
 
\section{Shear viscosity}
\label{chVI}

The observable behavior of the elliptic flow suggests that matter created in $AA$ collisions behaves as a near-perfect fluid with a very low viscosity over entropy density ratio \cite{PHENIX:2004vcz,PHOBOS:2004zne,BRAHMS:2004adc,STAR:2005gfr,ALICE:2010suc,CMS:2012zex,ATLAS:2011ah,ALICE:2011ab,CMS:2013bza}. It was proposed the indirect measurement of the shear viscosity over entropy density as a probe of the viscosity of the medium created in the collision. More recently, this probe has shown signs of a strongly interacting medium in small collision systems as well \cite{CMS:2010ifv,ALICE:2012eyl,CMS:2012qk,CMS:2016fnw,ATLAS:2017hap}. 

Assuming a simple kinetic model of an ideal gas of partons, it is possible to estimate the transport coefficients not in thermodynamic equilibrium, considering that the medium expands as a function of the initial state properties, as was initially proposed in \cite{DiasdeDeus:2011yk}. 

\begin{figure}[ht!]
 \begin{center}
  \includegraphics[scale=0.45]{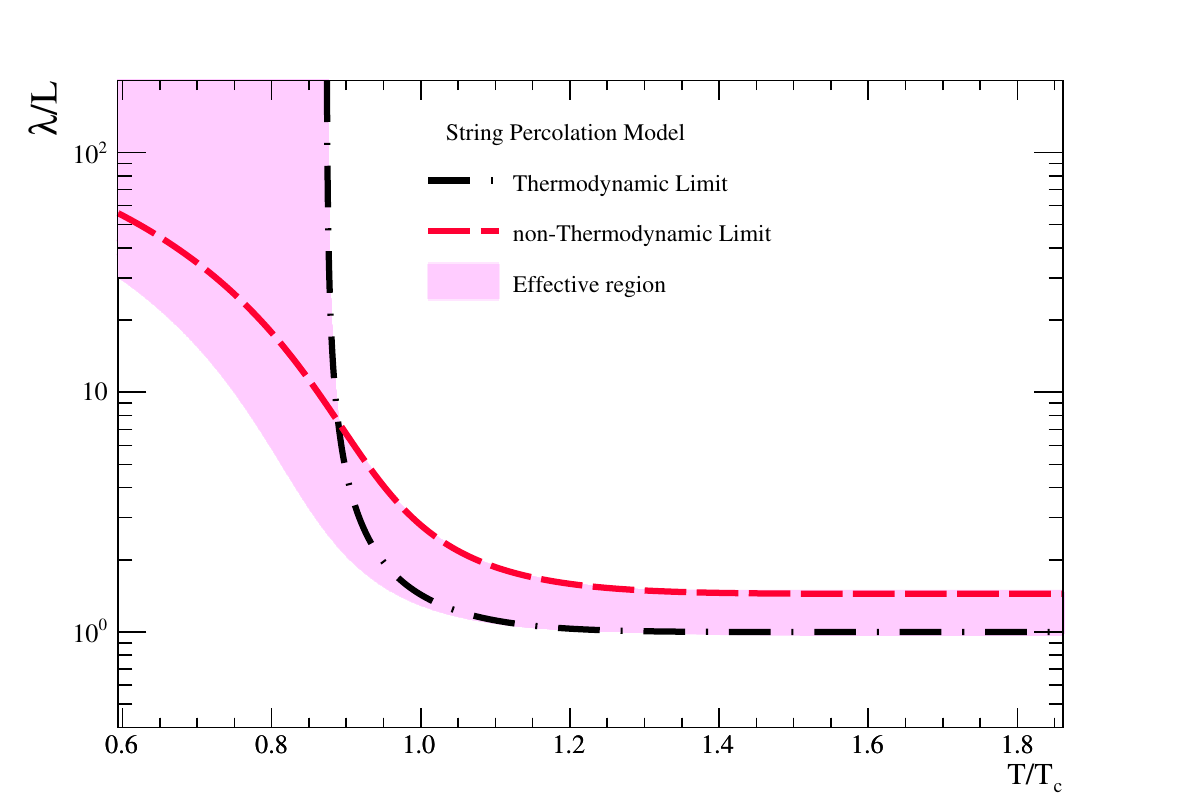} 
 \includegraphics[scale=0.45]{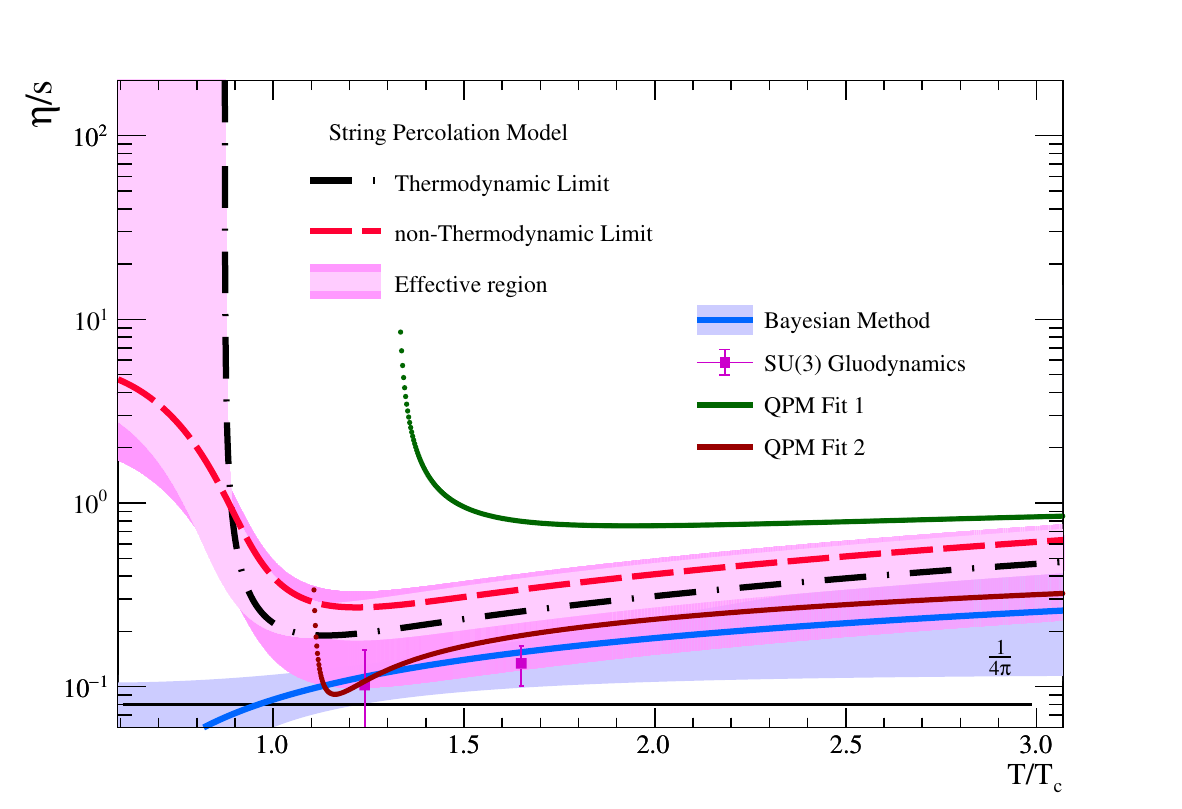}
\caption{
The (upper figure) mean free path and (lower figure) ratio of shear viscosity over entropy density as a function of $T/T_c$ both calculated on the SPM framework. The red dashed line corresponds to the nonTL computation, while the black dotted-dashed line is the TL one as reported in \cite{Bautista:2015kwa}, the pink area is the estimated effective region, and the magenta in $\eta/s$ is the extended effective region considering different $L$-size of sources.
The results from the Bayesian Method \cite{Bernhard:2019bmu} are shown in the blue region. 
The dark magenta squares correspond to SU(3) gauge calculations \cite{Meyer:2007ic} while the green and red solid lines correspond to QPM fits
\cite{Bluhm:2010qf}. The limit of AdS/CFT \cite{Policastro:2001yc} is included in the black continuum line.}
\label{fShear}
\end{center}
\end{figure} 

For computing the ratio of shear viscosity over entropy density $\eta /s$ in terms of the SPM parameters, we considered the relation given by the relativistic kinetic theory \cite{Hirano:2005wx}, which was also previously used for small collision systems \cite{Bautista:2015kwa,Gutay:2015cba}: 


\begin{equation}
    \frac{\eta}{s} = \frac{T\lambda}{5},
\label{eq.shear}
\end{equation}
where $\lambda = 1/(n\sigma_{tr})$ is the mean free path, with $n$ the number density of an ideal gas of partons and $\sigma_{tr}$ the transport cross section of its constituents. The number density is directly extracted from the initial number of strings damped by $F_s(\xi)$, which gives the collective medium effects:
\begin{equation}
    n = \frac{N_s F_s(\xi)}{SL},
\end{equation}
where $L \sim$ 1 fm represents the size that a string takes when extended in the beam axis direction. It is worth mentioning that this length has been taken as a first approximation, given that the calculations suggest that this value is in between $0.37$ fm and $1.2$ fm \cite{Baker:2019gsi}. On the other hand, we are considering that this formulation takes into account properties of the initial state, which is characterized by its corresponding 2-dimensional percolating system, with $N_s$ and $S$ as defined in Sec. \ref{chII}. In the same way, the transport cross section is given by $\sigma_{tr} = S_0 F_s(\xi)$, the transverse size of a single string multiplied by $F_s(\xi)$ \cite{DiasdeDeus:2011yk}. This leads us to estimate the mean free path using the definition of the string density from Eq.~\eqref{eq.stringd}: 
\begin{equation}
\lambda = \frac{L}{\xi F_s^2},
\label{sheareq}
\end{equation}
where $\xi F_s^2$ is the area covered by color sources, which is $1 - e^{-\xi}$ in the case of TL \cite{Ramirez:2019ekt}. 
Although the kinetic theory is applicable when the system is in equilibrium, in this case, we consider that a cluster is locally in equilibrium in the initial state for the appropriate validity of Eq.~\eqref{sheareq}. Moreover, the behavior of $\lambda/L$ is obtained for the complete system.
The upper Fig.~\ref{fShear} shows the $\lambda/L$ as a function of temperature. For the TL case, $\lambda$ decreases its value around 1 after the critical temperature \cite{DiasdeDeus:2011yk}. This scenario replicates a situation where particle movement is constrained by the effects of a strongly interacting medium, which aligns with what was observed in $AA$ collisions \cite{PHENIX:2004vcz,PHOBOS:2004zne,BRAHMS:2004adc,STAR:2005gfr}. Contrary to $pp$ collisions, where the medium does not achieve thermalization and its effects are less pronounced. Nevertheless, a reduction is still observed after the critical temperature region in the nonTL case. 


The behavior of $\lambda$ is inherited in the $\eta/s$ ratio, where a significant difference between the medium formed in heavy ion collisions and that from small collision systems is shown. The lower Fig.~\ref{fShear} shows a decrease in the $\eta/s$ ratio, which leads to an enhancement of collectivity effects. The results show that these effects are smaller in $pp$ collisions compared to those estimated for $AA$ collisions. The results on $\eta/s$ show that its minimum value for TL associated with $F(\xi)$ is reached at $T/T_c = 1.13187$, while the one associated with $F_s(\xi)$ for nonTL is reached at $T/T_c = 1.22508$; the results show an increase in the minimum value of $\eta/s$ by a factor of 1.4218 from $0.190018$ for TL to $0.270179$ for nonTL as shown in Fig.~\ref{fShear}. These results are compared with the fits for quasiparticle excitations with medium-dependent self-energies (QPM) \cite{Bluhm:2010qf} and the Lattice calculation in SU(3) gauge theory \cite{Meyer:2007ic}. The results from the Bayesian Method applied to heavy ion collisions \cite{Bernhard:2019bmu} are in between our estimation and, for $T>T_c$, the conjectured limit of AdS/CFT, $\eta/s \geq 1/(4\pi)$ \cite{Policastro:2001yc}.  
 
\begin{figure}[ht!]
\begin{center}
\includegraphics[scale=0.45]{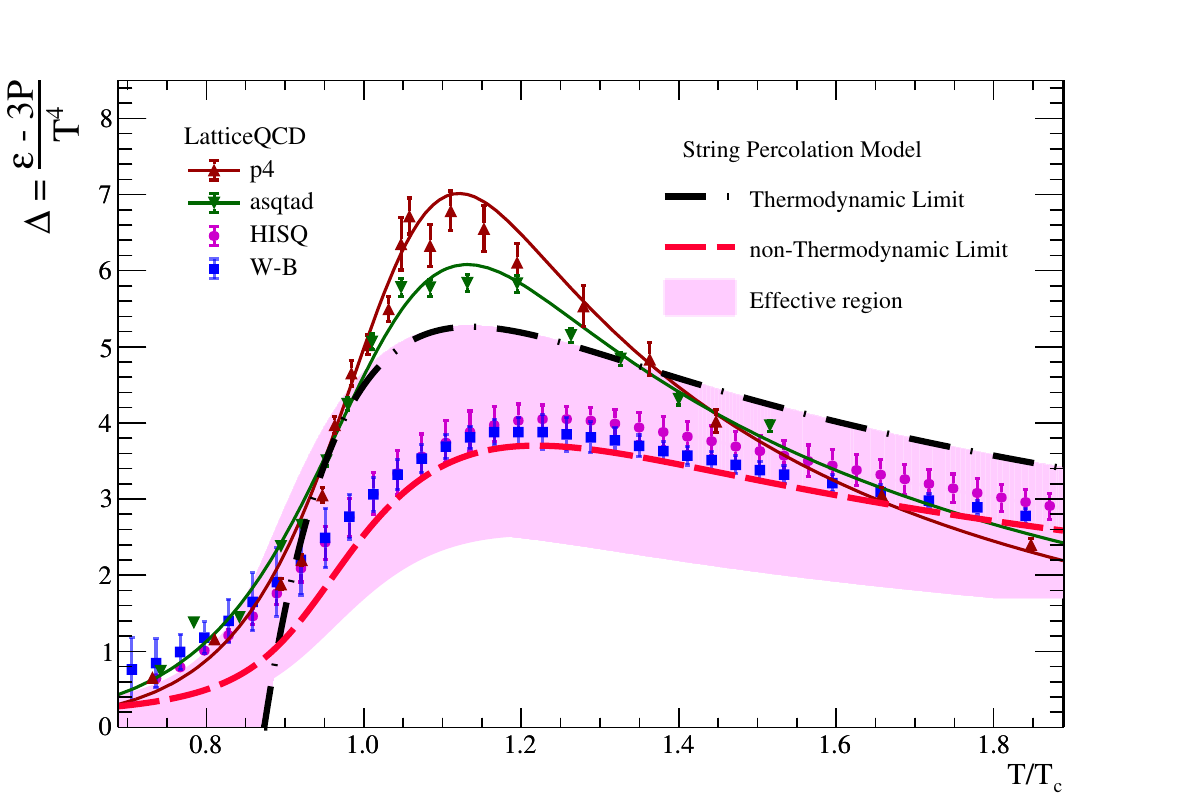}
\caption{
The behavior of the trace anomaly with respect to $T/T_c$ in TL (black dotted-dashed line) and nonTL (red dashed line) limits and the effective region (pink area) in the SPM framework compare with the results from Lattice QCD p4 (maroon triangles), asqtad (green triangles) \cite{Bazavov:2009zn}, HISQ (magenta circles) \cite{HotQCD:2014kol} actions and the Wuppertal-Budapest Collaboration (W-B) results (blue squares) \cite{Borsanyi:2013bia}.
}
\label{f1b}
\end{center}
\end{figure}

Trace anomaly $\Delta$ measures the deviation with respect to the conformal behavior and identifies residual interactions in the medium formed \cite{Gasbarro:2019kgj,Ishikawa:2013iia,Cheng:2009zi}. It is expected that this observable is related to the medium's viscosity properties. In previous works, it has been observed qualitatively that the trace anomaly can be approximated as the inverse of shear viscosity over entropy density \cite{Srivastava:2014nxa,DiasdeDeus:2016itj}:
\begin{equation}
\Delta \equiv \frac{\varepsilon - 3P}{T^4} \simeq \frac{s}{\eta_s}.
\label{eqDelta}
\end{equation}

Trace anomaly as well as the viscosity coefficients are susceptible to QGP phase transition \cite{Sch_fer_2009}. 
The behavior of the trace anomaly for nonTL goes accordingly to that reported by Wuppertal-Budapest Collaboration (W-B) using the Symanzik improved gauge and a stout-link improved staggered fermion action \cite{Borsanyi:2013bia}. And with the continuum extrapolated results from the HotQCD Collaboration of highly improved staggered quark action \cite{HotQCD:2014kol}. All of these results show a maximum value, for the SPM this maximum is located at $T/T_c = 1.13621$ in TL and at $T/T_c = 1.21918$ for nonTL (Fig.~\ref{f1b}).

\begin{figure}[ht!]
\begin{center}
\includegraphics[scale=0.45]{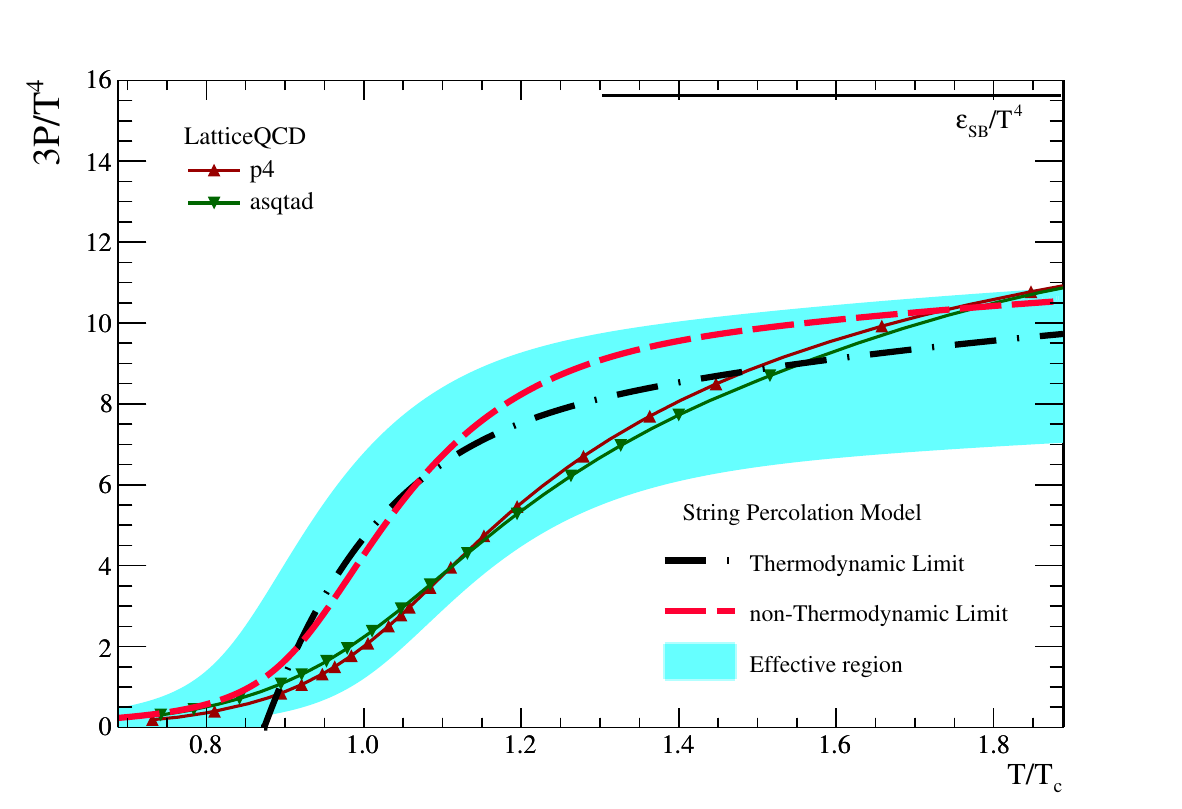}
\vspace{-10pt}
\includegraphics[scale=0.45]{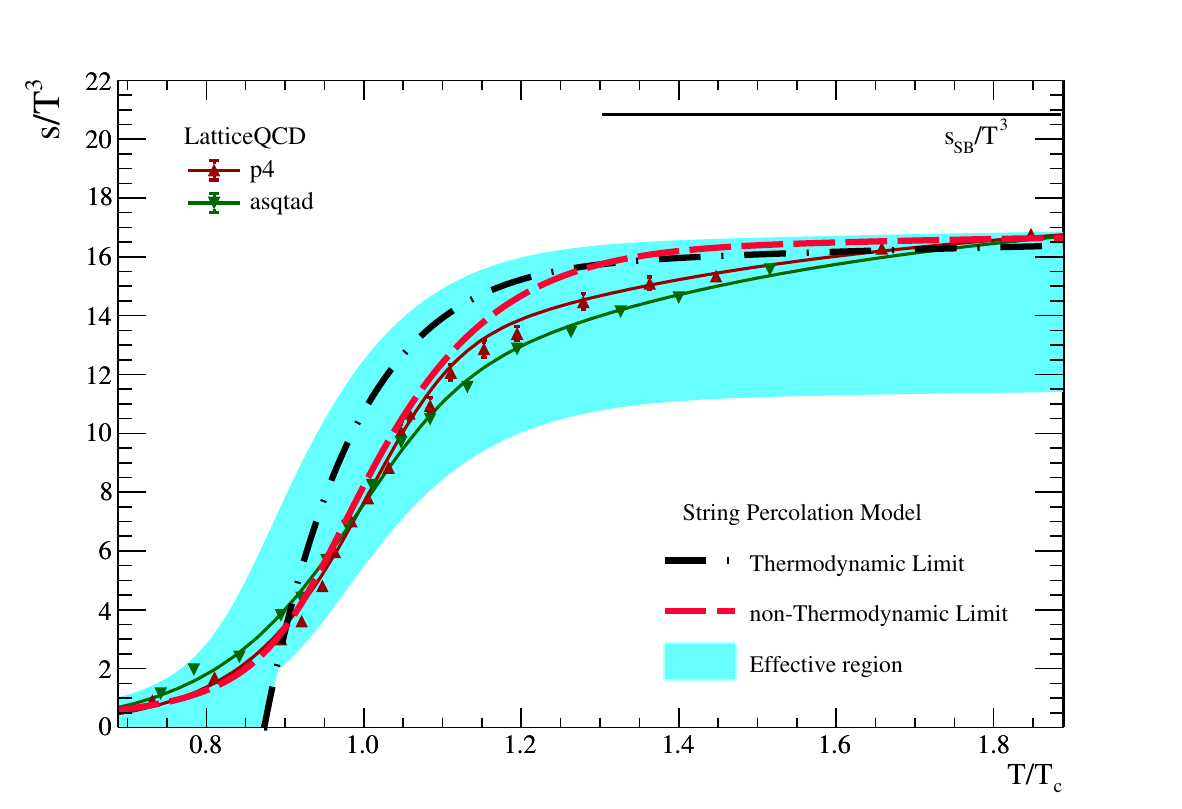}
\caption{ Behavior of the (upper figure) pressure over $T^4$  and (lower figure) entropy density over $T^3$ against $T/T_c$, for TL (black dotted-dashed lines) and nonTL (red dashed lines) cases, we show the effective estimate region in the blue area. In both figures, the comparison with Lattice QCD results p4 (maroon triangles) and asqtad (green triangles) actions with their respective parametrization (continuum lines) \cite{Bazavov:2009zn}
are included. }
\label{fs}
\end{center}
\end{figure}

Pressure $P$ is obtained from Eq.~\eqref{eqDelta} and from the first law of thermodynamics ($Ts =\varepsilon+ P $ \cite{Bjorken:1982qr});
we calculated the entropy density $s$ of the system. The results of $3P/T^4$ and $s/T^3$ are respectively shown in Fig.~\ref{fs} compared with LQCD  \cite{Bazavov:2009zn}. The SPM results agree with those of LQCD. Pressure begins saturation over 3$T_c$. For this reason, we can see the less pronounced slope in the decreasing region of trace anomaly after Fig.~\ref{f1b} maximal point.

\section{Bulk viscosity}
\label{chVII}

The effects of bulk viscosity are known to be very small, which in most of the high energy collisions were neglected due to the thermalization of the system. However, great attempts have been made to obtain their value from nuclear collisions from RHIC to  LHC \cite{Roy:2012np,Parkkila:2021tqq}.

Bulk viscosity corresponds to the resistance to the expansion of the fluid. The radial components seem damped due to the non-zero effect of bulk viscous pressure affecting the energy density profile of the created medium and converting it into pressure gradients changing the speed of sound $c_s^2$ \cite{Sahoo:2017umy}. This effect is related to the small perturbations produced in the medium formed \cite{Sahoo:2017umy}, such as vibrations and rotations of the medium components. In the SPM framework, these effects correspond to the fluctuations of string properties (color field, string tension, etc). 
To determine the bulk viscosity we calculate the speed of sound, which is given by a thermodynamic relation \cite{Bjorken:1982qr}:
\begin{equation}
c_s^2 = \left( \frac{\partial P}{\partial \varepsilon} \right)_s
= s \left( \frac{\partial T}{\partial \varepsilon} \right)_s
= -\frac{sT}{2\varsigma F_s}\cdot \frac{dF_s}{d\xi}  .
\label{e13}\end{equation}

\begin{figure}[ht!]
\begin{center}
\includegraphics[scale=0.45]{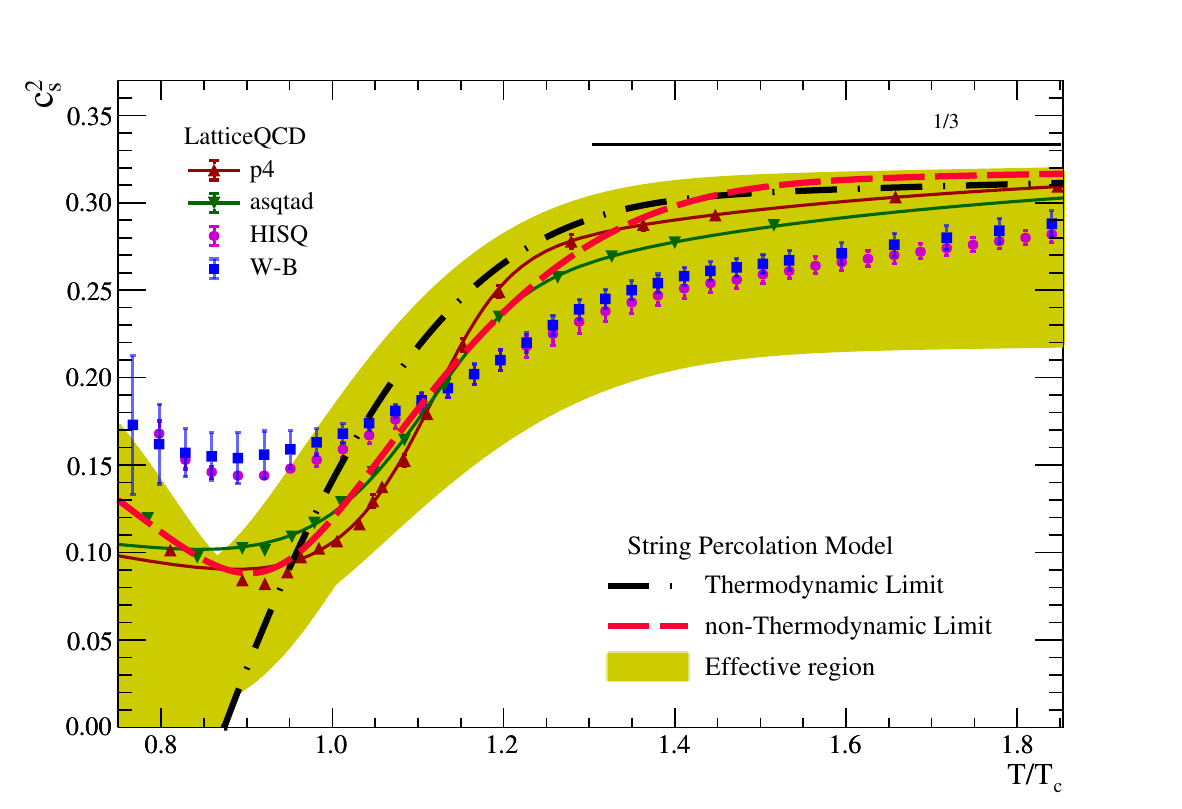}
\caption{Dependence of speed of sound squared with $T/T_c$ calculated in the SPM framework using Eq.~\eqref{e13}, the effective estimate region (golden area), the nonTL (red dashed line) and TL (black dotted-dashed line) limits are shown. The Lattice QCD results from W-B (blue squares)\cite{Borsanyi:2013bia}, the magenta circles are the HIQS action extrapolated results from the HotQCD Collaboration \cite{HotQCD:2014kol} p4 (maroon triangles) and asqtad (green triangle) \cite{Bazavov:2009zn} actions from HotQCD collaboration are compared.
}
\label{f5}
\end{center}\end{figure}

From Eqs.~\eqref{eq6} and \eqref{eq7} it is simple to obtain $dF_s/d\xi$ of Eq.~\eqref{e13}. In Fig.~\ref{f5} we compute the effective region of $c_s^2$. $F_s(\xi)$ gives a different behavior from what was previously reported in \cite{Braun:2015eoa,Sahoo:2017umy,Srivastava:2011vz,Scharenberg:2010zz}, 
and shows deviations from the results reported for elliptical geometry \cite{Ramirez:2019ekt} that are all below our parametrization.

Specifically, we observed a large deviation from TL in the region below the critical temperature, showing a ``dip-and-bump" effect, this behavior is in agreement with other phenomenological models \cite{He:2022kbc}, and goes accordingly with those reported from the Lattice QCD 2+1 flavor staggered fermion actions p4 and asqtad from \cite{Bazavov:2009zn}, the stout-link improved staggered fermion action from Wuppertal-Budapest Collaboration \cite{Borsanyi:2013bia} and the highly improved staggered quark action from HotQCD Collaboration \cite{HotQCD:2014kol}. 

\begin{figure}[ht!]
 \begin{center}
  \includegraphics[scale=0.45]{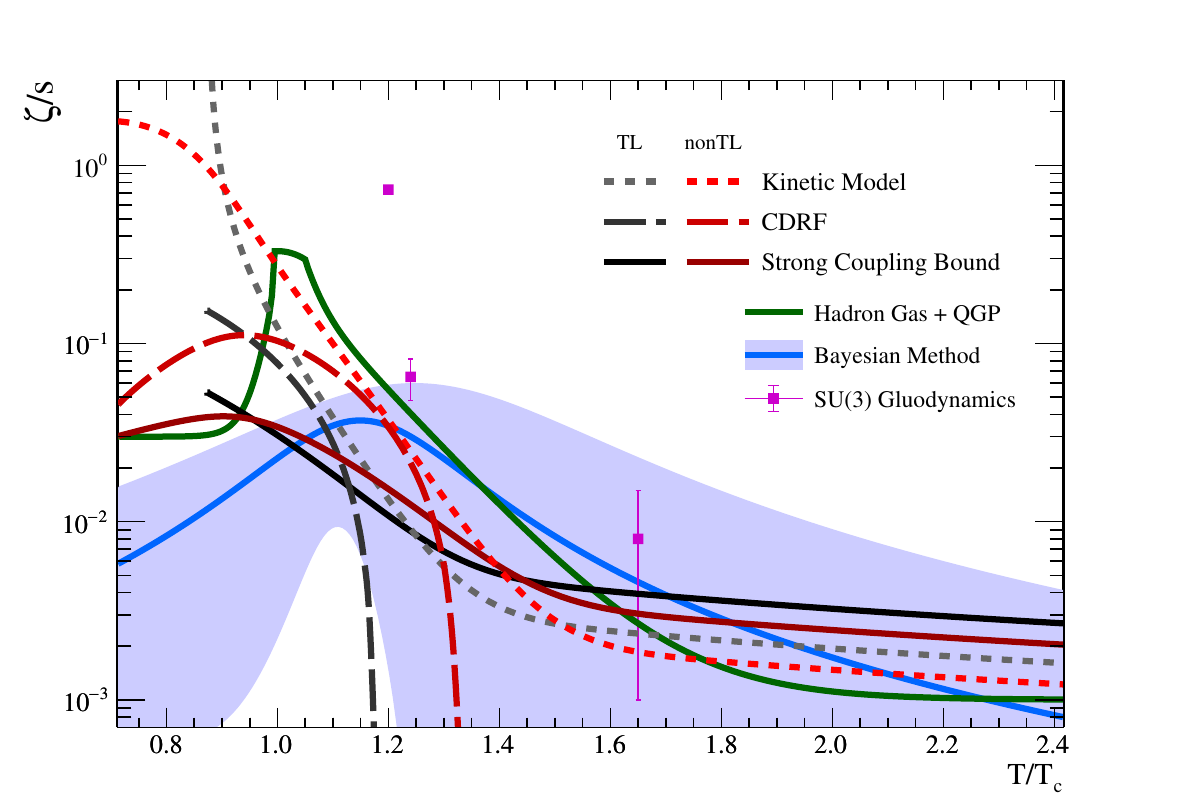}
\caption{Bulk viscosity over entropy density as a function of $T/T_C$ from the kinetic model Eq.~\eqref{Bkinetic} (dotted lines), the CDRF formalism Eq.~\eqref{bulkeq} (dashed lines), and the holographic limit using Eq.~\eqref{SCbulk} (continuum lines) calculated on the SPM framework are shown, distinguish TL in gray scale lines and nonTL in red lines. The Bayesian Method results \cite{Bernhard:2019bmu} (blue region), the parametrization of Hadron Gas to QGP (continuum green line), and the SU(3) gauge theory calculations \cite{Meyer:2007dy} are included.}
\label{fBulk}
\end{center}
\end{figure}

The first approximation of bulk viscosity over entropy density $\zeta/s$ of the simplest kinetic model in classical statistics with relaxation time approximation is given by \cite{Dusling:2011fd}:
\begin{equation}
    \frac{\zeta}{s} = 15\frac{\eta}{s}\left( \frac{1}{3} - c_s^2 \right)^{2},
    \label{Bkinetic}
\end{equation}
which describes the bulk viscosity coefficient in terms of shear viscosity and speed of sound calculated in the SPM framework. The result of this approach using $F(\xi)$ and $F_s(\xi)$ is shown in dashed lines of Fig.~\ref{fBulk}. This approximation exhibits a monotonically decreasing behavior.


As a second approach, we use the results reported and verified in \cite{Koide:2007ja,Koide:2008nw,Koide:2009sy,Huang:2010sa,Denicol:2010br,Koide:2005qb} of the projection operator approach to obtain the microscopic formulas for the transport coefficients in causal dissipative relativistic fluid dynamics (CDRF), in terms of the SPM observables $T$, $s$, $\Delta$, $c_s$.
The reported microscopic formula of the bulk viscosity $\zeta$ with its respective relaxation time $\tau_\Pi$ of CDRF is given by \cite{Huang:2011ez,Koide:2007ja,Koide:2008nw,Koide:2009sy,Huang:2010sa,Denicol:2010br,Koide:2005qb}:
\begin{equation}
\frac{\zeta}{s} = \left( \frac{1}{3} - c_s^2 \right) \tau_{\Pi} T - \frac{2 \tau_{\Pi} T^4}{9s}\Delta ,
\label{bulkeq}
\end{equation}
where $\tau_\Pi$ is considered of the order of a fermi and the fraction $ 2/9 $ has to do with the number of fermionic degrees of freedom. 

The results of Eq.~\eqref{bulkeq} in TL and nonTL are shown in Fig.~\ref{fBulk} labeled as CDRF. In which there are differences in their behaviors and shifts in their vanishing points are shown.

On the other hand, in \cite{Buchel:2007mf} is conjectured a lower limit on bulk viscosity of strongly coupled gauge theory plasmas as:
\begin{equation}
    \frac{\zeta}{s} \geq   2\left( \frac{1}{3} - c_s^2 \right) \frac{\eta}{s} \geq \frac{1}{2\pi}\left( \frac{1}{3} - c_s^2 \right),
\label{SCbulk}
\end{equation}
considering $\eta/s \geq 1/(4\pi)$ \cite{Policastro:2001yc}. Which is shown in solid black (TL) and red (nonTL) lines in Fig.~\ref{fBulk}.

Fig.~\ref{fBulk} includes the results of $\zeta/s$ reported from viscous relativistic hydrodynamics Bayesian Method \cite{Bernhard:2019bmu}, that is near the holographic limit \cite{Buchel:2007mf} and CDRF calculations in the SPM framework. 

We compare our results of $\zeta/s$ with the hadron resonance gas model, which incorporates all identified particles and resonances with masses below 2 GeV, along with a density of Hagedorn states that increases exponentially for masses exceeding 2 GeV \cite{Noronha-Hostler:2008kkf}, and 
the continuum parametrization of the LQCD equation of state results of  $\zeta/s$ of hot quark-gluon matter in the presence of light quarks \cite{Karsch:2007jc}, as presented in \cite{Denicol:2009am,Ryu:2015vwa}. These models show a similar behavior to the kinetic model considered in the SPM ($T > T_c$ region). Also, the Lattice gluodynamics calculation in SU(3) gauge theory \cite{Meyer:2007dy} shows a scaled same dependence.

\begin{table}[ht!]
    \centering
        \caption{Results of the maximum value of bulk viscosity over entropy density for different approaches in TL and in our parametrization (nonTL).}
    \begin{ruledtabular}
    \centering    
    \begin{tabular}{ c c c c c } 
	&	\multicolumn{2}{c}{TL}		&	\multicolumn{2}{c}{nonTL}		\\ \hline
	&	$\zeta/s$ (Max)	&	$T/T_c$	&	
$\zeta/s$ (Max)	&	$T/T_c$	\\ \hline
Kinetic	&	35132.6	&	0.873553	&	1.84112	&	0.146509	\\  
CDRF	&	0.152694	&	0.873553	&	0.111427	&	0.940129	\\ 
Bound	&	0.0530515	&	0.873553	&	0.0390168	&	0.903653	\\ 
             \end{tabular}
    \end{ruledtabular}   
    \label{tab:bulkmax}
\end{table}

\begin{figure}[ht!]
    \centering
    \includegraphics[scale=0.45]{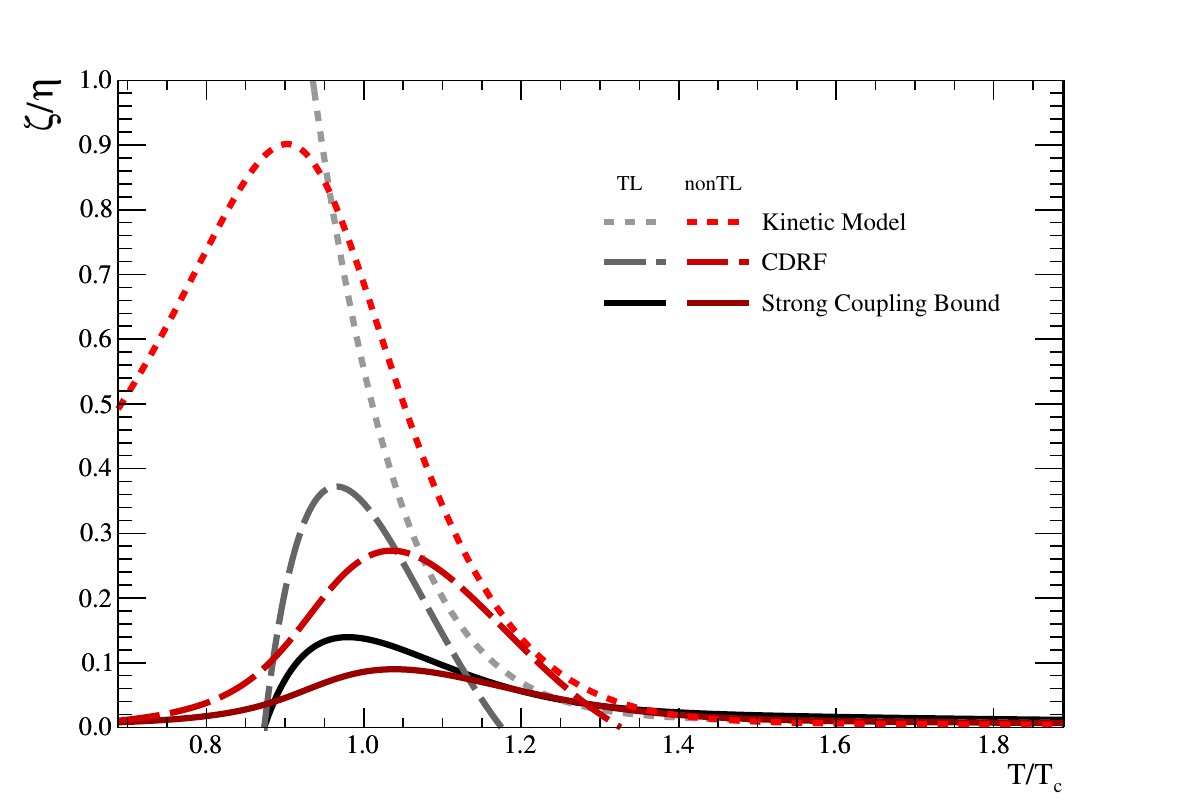}
    \caption{The results of the bulk-shear viscosity ratio against $T/T_c$ from the kinetic theory (dotted lines), CDRF (dashed lines), and the gauge theory plasma at strong coupling limit (continuum lines) \cite{Buchel:2007mf} 
    calculated in the SPM framework. Distinguish TL in gray lines and nonTL in red lines. }
    \label{fResult}
\end{figure}

In Table \ref{tab:bulkmax}, we present the maximum values of $\zeta/s$ for the kinetic theory, CDRF formalism, and conjectured bound with its respective $T/T_c$ value for TL and nonTL. We can observe that in all cases $\zeta/s$ reaches its maximum value below the critical temperature. And the TL goes much higher than nonTL, reaching its maximum value. For the CDRF formalism, it is reached just $73.54\%$ of the TL. For the lower conjectured bound it is $72.97\%$ and for the case of the simplest kinetic model, we see a discrepancy in the values around $0.0052\%$, because the value of $c_s^2$ for TL vanishes at $T/T_c = 0.873553$, Eq.~\eqref{Bkinetic} gives the same tendency as shear viscosity over entropy density in TL.

In Fig.~\ref{fResult} we show the interplay between shear and bulk viscosity given by the ratio $\zeta/\eta$  computed in the SPM framework, where we can observe a maximum value around the critical temperature region in all cases.
For the case of the Kinetic Model, the $\zeta/\eta$ ratio shows its maximum value in $T < T_c$ for TL and nonTL scenarios, and for the CDRF formalism we can see a change in the slope in the region just below $T=T_c$ for the TL case, and for nonTL right before $T = T_c$. 

\section{Conclusions}

We have computed the $\eta/s$ and $\zeta/s$ ratios by proposing a global parametrization for the SPM color reduction factor, that considers the small-bounded effects in the geometrical phase transition at the non-thermodynamic limit for $\mu_B=0$. 
Our description highlights the differences in the physics behind both TL and nonTL cases discussed in the framework of the SPM for the QCD matter formed in $pp$ and $AA$ collisions at LHC energies. 

The ratio $\eta/s$ is estimated in a simplified kinetic formulation of an ideal gas of partons. The results on this coefficient show different minimum values in TL and nonTL, within the region $T_c < T  < 1.23 T_c$. 
The first implication related to the nonTL case reveals that a phase transition must occur at higher temperatures
since a shift in the inflection point is found.
Additionally, the medium effects that constrains particle movement are less pronounced on this limit, as can be seen in Fig.~\ref {fShear}.

The behavior of the speed of sound shows that it does not vanish at low temperatures (Fig.~\ref{f5}). 
This is important since $c_s^2$ acquires relevance in the computation of $\zeta/s$, making its contribution non-neglectable. 

For the estimation of $\zeta/s$, we used two formalisms and compared them to a  conjectured strong coupling bound.  
By using the CDRF on the SPM, we found that $\zeta/s$ vanishes at $T/T_c = 1.17$ for TL and at $T/T_c = 1.32$ for nonTL, as shown in Fig.~\ref{fBulk}. On the other hand, the implementation of the simplest kinetic model guides $\zeta/s$ to reach higher values before critical temperature, as summarized in Table \ref{tab:bulkmax}. Both formalisms are above the strong coupling bound for $T/T_c < 1.1$, where the maximum values are reached. Hence, the contribution of $\zeta$ becomes relevant in this region, implying that there is a strong effect driven by the fluctuations of the string properties, such as the color field, as well as the string tension. 
In addition, the ratio $\zeta/\eta$ for nonTL scenarios shows a shift of the maximum point, which is reached at higher temperatures. This implies that nonTL requires a higher temperature in order to reach phase transition due to the fluctuations coming from the bulk contributions (Fig.~\ref{fResult}).


It is important to acknowledge certain limitations in our estimations of transport coefficients. First, our approach assumes that a cluster is locally in equilibrium in the initial state, which is crucial for the validity of Eq.~\eqref{sheareq}. However, deviations from local equilibrium, especially in the early stages of high-energy collisions, could impact the accuracy of our estimations.


Our results show a clear difference in the behavior of the observables that take into account the finite size effects with respect to those that are predicted in the thermodynamic limit, showing that the fluctuations of the initial state have a qualitative relevance in the estimation of the coefficients that characterize the medium formed in small collisions systems. 

While our study provides valuable insights into the behavior of transport coefficients in small collision systems within the framework of the String Percolation Model, one should remain mindful of these limitations when interpreting and applying our results.


\section*{Acknowledgements} 

We thank CONAHCyT-M\'exico for supporting this work under the projects CF-2019/2042, and A1-S-26507. On the other hand, J. R. A. G. and P. F. thank CONAHCyT-M\'exico for the graduate fellowships 645654 and 848955 respectively.

\bibliographystyle{elsarticle-num}

\bibliography{bib}

\end{document}